\documentclass[final,english]{bullsrsl}[]

\usepackage[latin1]{inputenc}
\usepackage[T1]{fontenc}

\usepackage{natbib} 
\usepackage{graphicx}

\begin{document}
\title{Optical spectroscopy of 1A~0535+262 before, during, and after the 2020 giant X-ray outburst}

\author[affil={1}, corresponding]{Sachindra}{Naik}
\author[affil={1,2}]{Birendra} {Chhotaray}
\author[affil={1}]{Neeraj}{Kumari}
\affiliation[1]{Physical Research Laboratory, Navrangpura, Ahmedabad - 380009, India}
\affiliation[2]{Indian Institute of Technology, Gandhinagar-382355, India}
\correspondance{snaik@prl.res.in}
\maketitle



\begin{abstract}
We present the findings from our study of the Be/X-ray binary 1A~0535+262/HD~245770 during the giant X-ray outburst in October 2020. We utilized the 1.2-m telescope at Mount Abu Infrared observatory for optical observations of the Be companion star. The outburst reached a peak X-ray flux of approximately 11 Crab in the 15-50 keV range, marking the highest ever recorded X-ray outburst from the pulsar. We conducted optical observations in the 6000-7200 \AA~ range before, during, and after the X-ray outburst, aiming to examine the evolution of the circumstellar disc of the Be star from February 2020 to February 2022. Our optical spectra displayed prominent emission lines at 6563 \AA ~(H I), 6678 \AA ~(He I), and 7065 \AA ~(He I). Notably, the H$\alpha$ line exhibited significant variability in the spectra. Prior to and during the outburst, the line profiles appeared single-peaked, and asymmetric with broad red and blue wings, respectively. However, post-outburst observations revealed a double-peaked profile with asymmetry in the blue wing. Our pre-outburst observations confirmed a larger Be circumstellar disc that diminished in size as the outburst progressed. Additionally, the observed variations in the H$\alpha$ line profile and parameters indicate the presence of a highly misaligned, precessing, and warped Be disc.
\end{abstract}


\section{Introduction}

Be/X-ray binaries (BeXRBs) are considered to be the largest sub-group of high mass X-ray binaries (HMXBs). These systems primarily consist of a neutron star and a massive non-supergiant O and B type stars (\citealt{ziolkowski2002x, paul2011transient, reig2011x}) orbiting around the common center of mass. The neutron stars in these binaries are fueled by mass accretion from the companion Be star. In BeXRBs, the companion Be star is bright in optical and infrared wavebands. They also exhibit specific emission lines (H~I, He~I, and Fe~II) and an excess of infrared radiation~\citep{porter2003classical}. Unlike the classical O and B-type stars, the Be stars have these unique characteristics due to the presence of an equatorial circumstellar disc formed by material discharged from the rapidly rotating Be star near its critical velocity~\citep{porter2003classical}.

Most of the BeXRBs have wide and eccentric orbits (e$\geq$0.3) with orbital periods of $\geq$ 10 days. Mass transfer from the Be star to the compact object results in transient X-ray outbursts at the periastron passages. However, all BeXRBs are not transients \citep{1999MNRAS.306..100R}. The transient X-ray activities are of two types : normal (Type~I) and giant (Type~II) X-ray outbursts. Normal outbursts occur periodically near the periastron passage of the neutron star and last for about 20\% of the orbital period, reaching peak luminosities of $L_{\rm X}$ $\sim$ 10$^{35-37}$ erg s$^{-1}$. Giant outbursts, on the other hand, are irregular, rare and not influenced by orbital modulation. They last for several orbital periods and have peak luminosities of $L_{\rm X}$ $\geq$10$^{37}$ erg s$^{-1}$. The initiation of X-ray outbursts is generally attributed to mass transfer from the Be circumstellar disc to the neutron star, although the precise mechanism remains under investigation. \citet{okazaki2001natural} proposed that the normal X-ray outbursts in BeXRBs with medium to high eccentric orbits are explained by the resonantly truncated decretion disc model \citep{2001A&A...369..108N,10.1046/j.1365-8711.2002.05960.x}. Observationally it is also seen that the disc of the Be star in binaries are more denser than the isolated ones \citep{2001A&A...367..884Z, 2017MNRAS.471..596C}. \citet{10.1093/pasj/65.2.41} studied 1A~0535+262 and 4U~0115+634, revealing that the normal outbursts result from radiatively inefficient accretion flow (RIAF) from a tidally truncated Be disc to the neutron star. They suggest that the giant X-ray outburst occurs when the Be star's disc is misaligned with the binary orbit, causing disc warping and Bondi-Hoyle-Lyttleton (BHL) accretion onto the neutron star. \citet{Martin_2014_a}  also proposed that the giant outbursts originate from highly misaligned, eccentric, and warped Be disc. Warping episode of discs were observed in systems such as $\gamma$~Cas, 59~Cyg, 4U 0115+63, and 1A 0535+262. \citep{1998A&A...330..243H} suggested an inclined Be disc for  $\gamma$~Cas and 59~Cyg by studying the emission line variations from shell to single-peaked profile. \citet{2007A&A...462.1081R}  and \citet{2018A&A...619A..19R} also suggested the presence of warped disc during the giant outbursts in 4U~0115+63. 

1A~0535+262 is an active BeXRB transient with frequent Type I X-ray outbursts. It was discovered by the Ariel~V space telescope in 1975 during a giant X-ray outburst \citep{rosenberg1975observations}. It consists of a neutron star orbiting an O9.7 IIIe star \citep{giangrande1980optical} in a relatively wide and eccentric orbit (eccentricity $e$ $\sim$0.47) with an orbital period of approximately 110 days \citep{finger1994hard}. The spin period of the neutron star was estimated to be around 104 s \citep{rosenberg1975observations}. Since its discovery, 1A~0535+262 has undergone several giant X-ray outbursts (\citet{Camero_Arranz_2012} and references therein). \citet{10.1046/j.1365-8711.1999.02112.x} found that the outburst in 1A~0535+262 occurred during the optical fading phase of the Be star. Spectroscopic observations of 1A~0535+262/V725~Tau revealed a strong H$\alpha$ line variability \citep{10.1093/pasj/63.4.L25}. Before the giant X-ray outburst, the H$\alpha$ line was observed to evolve from various shapes to a single peak with increasing strength and Full-Width at Zero Intensity (FWZI) \citep{1998MNRAS.294..165C,10.1111/j.1365-2966.2006.10127.x, Camero_Arranz_2012}. This behavior of the H$\alpha$ line is attributed to the warping of the Be companion's circumstellar disc prior to the giant outburst. \citet{article_Moritani}, using high dispersion optical spectroscopic observations during the 2009 and 2011 giant X-ray outbursts, suggested that a precessing warped Be disc triggered the giant X-ray outburst in 1A~0535+262. 

An extensive near-infrared spectroscopic and photometric observations of the Be/X-ray binary 1A~0535+262/HDE~245770 during the 2011 February-March giant X-ray outburst revealed a gradual and systematic fading in the J, H, and K-band light curves since the onset of the X-ray outburst \citep{2012RAA....12..177N}. This suggested a mild evacuation/truncation of the circumstellar disk of the companion Be star during the outburst. Near-infrared spectroscopy of the object showed that the J, H, and K-band spectra are dominated by the emission lines of hydrogen Brackett and Paschen series and He~I lines at 1.0830, 1.7002 and 2.0585 $\mu$m. The presence of all the hydrogen emission lines in the near-infrared spectra, along with the absence of any significant change in the continuum of the Be companion star during the X-ray quiescent and X-ray outburst phases, suggested that the near-infrared line emitting regions of the disk are not significantly affected during the 2011 giant X-ray outburst \citep{2012RAA....12..177N}. The pulsar in 1A~0535+262/HDE~245770 system went into an intense giant X-ray outburst in October 2020. In order to have a clear understanding of the effect of the giant X-ray outburst on the circumstellar disc of the Be star, we carried out optical spectroscopic observations of the system. In Section~2, we present the optical observations and data reduction procedures adopted in the present work. In Section~3, the spectroscopic results are presented, followed by the discussion and conclusion section. 

\section{Observations and Data Reduction}
The X-ray outbursts observed in BeXRB systems are the result of the abrupt accretion of a significant amount of matter from the circumstellar disc of the Be star onto the neutron star. Understanding the X-ray properties of the neutron star during these outbursts is closely linked to changes in the properties of the Be star. Therefore, conducting optical/infrared observations of BeXRBs during X-ray outbursts is of great importance to understand the physical condition of the Be disc. As a part of the ongoing project, optical spectroscopic observations of BeXRB 1A~0535+262 were carried out using the 1.2 m telescope at Mount Abu Infrared Observatory (MIRO), utilizing the Faint Object Spectrograph and Camera-pathfinder (MFOSC-P) instrument \citep{2018SPIE10702E..4IS}. The instrument offers seeing-limited imaging capabilities in Bessel B, V, R, and I filters, with field-of-view of 5.2$^{'}\times$5.2$^{'}$  and a sampling rate of 3.3 pixels per arcsecond. The spectroscopic observations were conducted using the 75 $\mu$m  slit, equivalent to 1$^{''}$ on the sky, in R2000 mode, the highest resolution available for the MFOSC-P instrument. The R2000 resolution mode provides a spectral coverage of $\sim$6000-7200~\AA. The raw data of the MFOSC-P observations were processed using the custom data analysis routines in python, incorporating open-source image processing libraries such as ASTROPY (see \citealt{2022MNRAS.510.4265K}). The reduction steps involved bias subtraction, cosmic ray removal, sky subtraction, wavelength calibration using Neon and Xenon calibration lamps, instrument response correction using the standard star SAO~77466, and generation of science images. 

\begin{figure}
\vspace*{-2.3cm}
\centering
\includegraphics[scale=0.44]{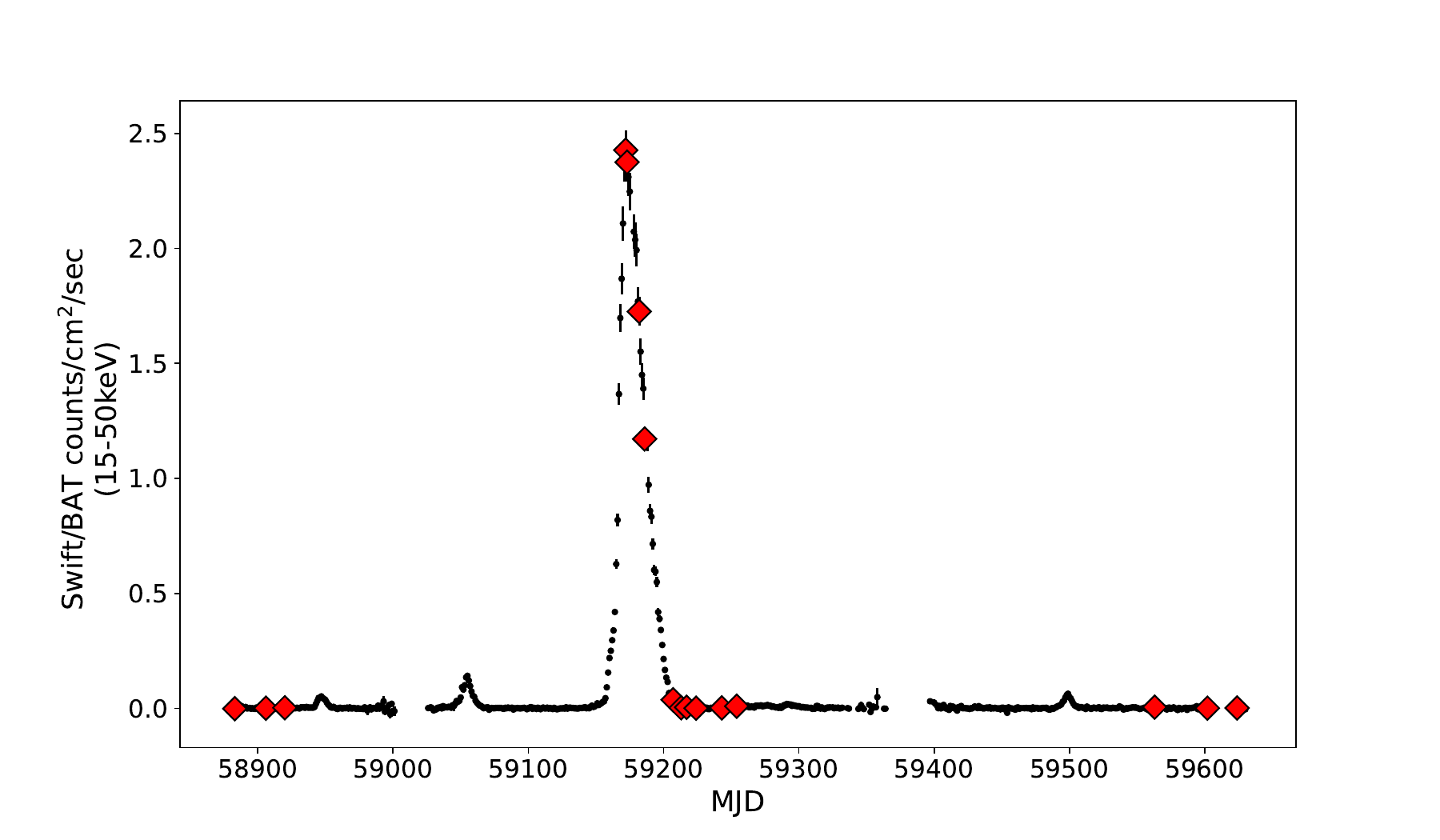}
\bigskip

\begin{minipage}{12cm}
\caption{Swift/BAT daily light curve (15-50 keV) of pulsar 1A~0535+262 during the 2020 giant X-ray outburst (black points). Red diamonds represent the epochs of optical observations of the binary with the 1.2 m telescope at Mount Abu using the MFOSC-P instrument. For reference: 1 Crab = 0.220 counts cm$^{-2}$ s$^{-1}$ in the 15-50 keV band.}
\label{fig:swiftlc}
\end{minipage}
\end{figure}

\section{Spectroscopic Results}

The spectroscopic observations were carried out at several epochs as shown in Figure~\ref{fig:swiftlc}. In Figure~\ref{fig:swiftlc}, the optical observations are plotted simultaneously with the Swift/BAT light curve to show X-ray brightness of the neutron star at the time of observation. In this section, we present the optical spectroscopic results obtained before, during, and after the giant X-ray outburst. The spectroscopic observations of the BeXRB were done using MFOSC-P in R2000 mode. Initial observations were made on February 4, 27, and March 12, 2020, during the neutron star's quiescent state. These observations  show the properties of the Be circumstellar disc before the X-ray outburst. As the system went onto outburst, subsequent closely spaced observations were made from November 19, 2020, to February 9, 2021, during the giant X-ray outburst. Three more observations were made on December 15, 2021, January 23, 2022, and February 14, 2022, after the system return to the quiescent state.

\begin{figure*}
	\vspace*{-2.0cm}
	\centering
	\includegraphics[scale=0.45]{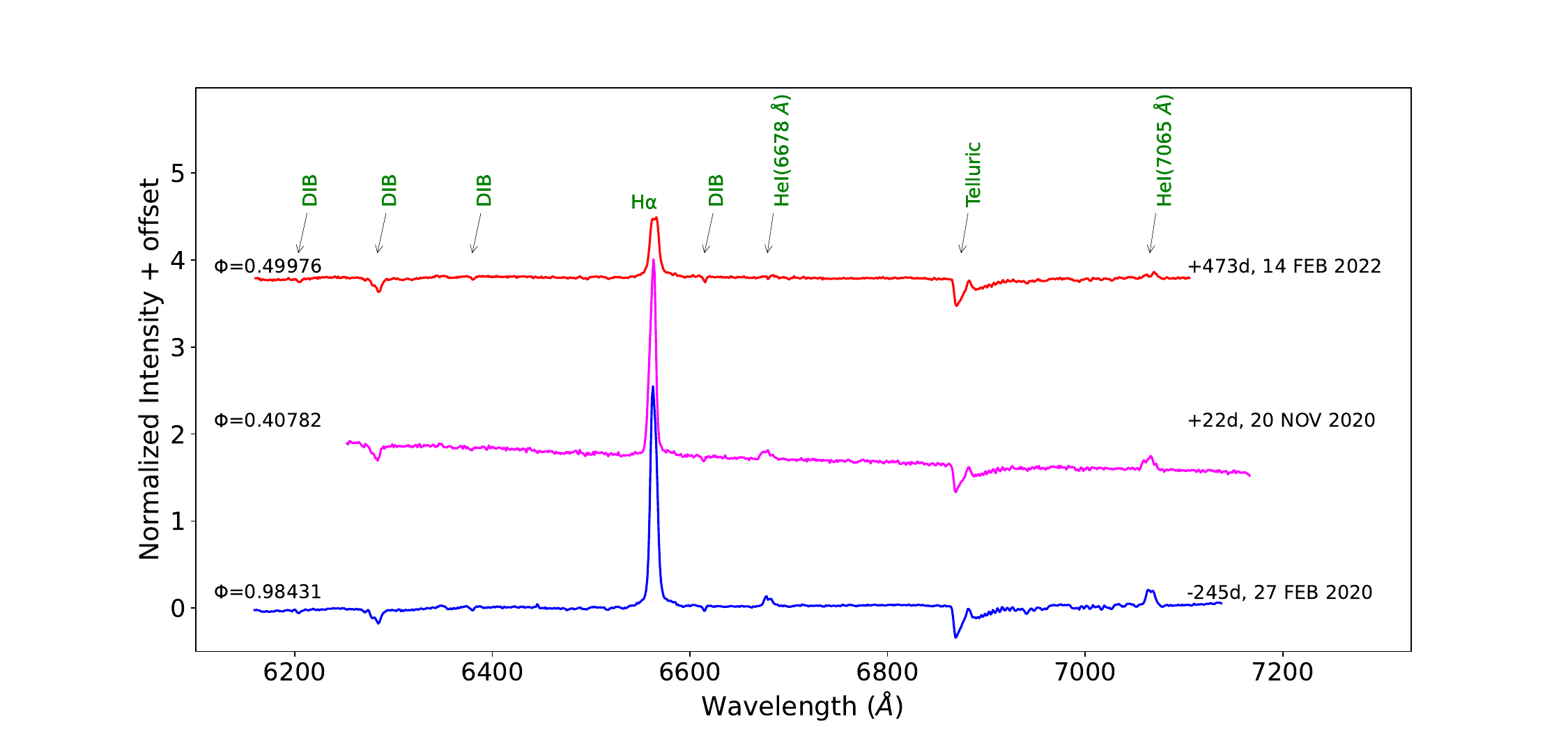}
	\bigskip
	\begin{minipage}{12cm}
	\caption{Evolution of optical spectrum of the Be/X-ray binary 1A~0535+262 using MFOSC-P instrument. Spectra are not reddening corrected. Observation dates and corresponding orbital phases ($\phi$) of the neutron star are annotated in the figure. Days before and after the onset of the giant X-ray outburst on MJD~59151 are indicated in -ve and +ve signs. Blue, magenta, and red colors correspond to the spectra before, during, and after the outburst, respectively.}
	\end{minipage}
	\label{fig:spec_evolve}
\end{figure*}

Figure~\ref{fig:spec_evolve} shows the 6000-7200 \AA~spectra obtained using MFOSC-P in R2000 mode for 1A~0535+262/HD~245770. The continuum normalized intensities are plotted with offsets for clarity. The observation date and days before/after the giant X-ray outburst are indicated in the figure. Here, the spectra of only 3 epochs are presented to show the overall picture before, during, and after the X-ray outburst. The spectra within a specific outburst phase (i.e., before, during and after the X-ray outburst) exhibit a similar continuum slope, and all the lines, including H$\alpha$ (6562.8 \AA), HeI (6678 \AA), and HeI (7065 \AA), are observed in emission. The overall shape of the emission lines also remains same in the spectra within a particular phase of the outburst. Considering this, in Figure~\ref{fig:spec_evolve}, we chose to present a single epoch spectrum for each phase (before, during, and after) of the giant X-ray outburst. Orbital phases ($\phi$) indicate the neutron star's position in the binary orbit during the observation. The phases are estimated by using the method provided in \citet{article_Moritani} and $\phi$=0 corresponds to the periastron passage of the neutron star. Diffuse Interstellar Bands (DIBs) and atmospheric telluric features are detected in the spectra of all epochs. Four DIBs are identified at specific wavelengths, such as 6203.06 \AA, 6283.86 \AA,  6379.20 \AA, and 6613.62 \AA~\citep{1995ARA&A..33...19H}. The telluric features, like the O2 B band at $\lambda$ $\sim$ 6870 \AA, are also present in the spectra.

\begin{figure}
	\vspace{-2.0cm}
	\centering
	\includegraphics[height=9cm, width=14cm]{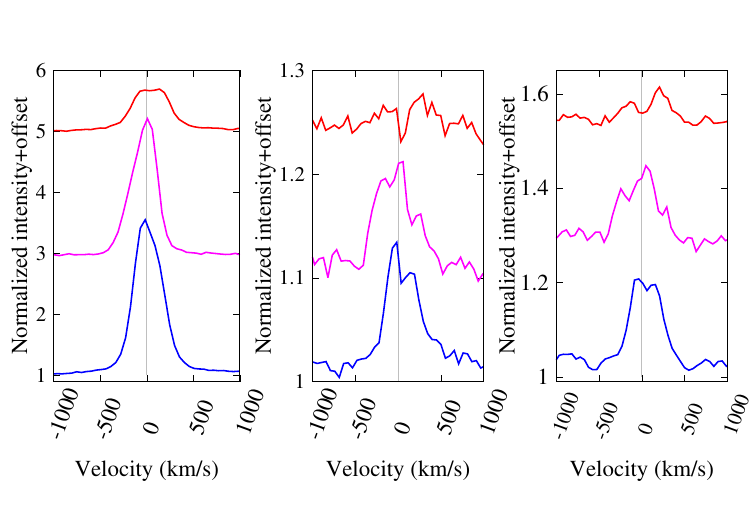}
	\bigskip
	\begin{minipage}{12cm}
	\caption{The H$\alpha$ (left), HeI (6678 \AA) (middle), and HeI (7065 \AA) (right)  line profiles of 1A~0535+262/HD~245770 during the pre-outburst (blue color), outburst (magenta color) and post-outburst (red color) observations. The errors on the flux values are of 1$\sigma$ level.  The profiles are plotted with certain offsets for clarity.}
	\label{fig:halpha}
	\end{minipage}
\end{figure}

Figure~\ref{fig:halpha}, show the evolution of H$\alpha$ (6562.8 \AA), HeI (6678 \AA), and HeI (7065 \AA) emission line profiles of spectra shown in Figure~\ref{fig:spec_evolve}. The X-axis represents velocity (km/s) to visualize emission from different parts of the circumstellar disc. Negative velocities indicate the blue-shifted components, while positive velocities represent the red-shifted components. During the pre-outburst phase, the H$\alpha$ lines are single-peaked, and moderately asymmetric with a broad red-wing, while HeI (6678 \AA) and HeI (7065 \AA) lines are double-peaked with different blue and red component intensities. During the X-ray outburst, the H$\alpha$ lines become asymmetric but with a broader blue component unlike pre-outburst phase, and HeI (6678 \AA) and HeI (7065 \AA) lines exhibit complex multi-peaked profiles. In the post-outburst phase, the H$\alpha$ lines evolved to double-peaked profile. The HeI (6678 \AA) and HeI (7065 \AA) lines show multiple components. In the last three observations, all lines become double-peaked with comparable peak intensities.

\begin{figure*}
	\vspace{-2.0cm}
	\centering
	\includegraphics[scale=0.25]{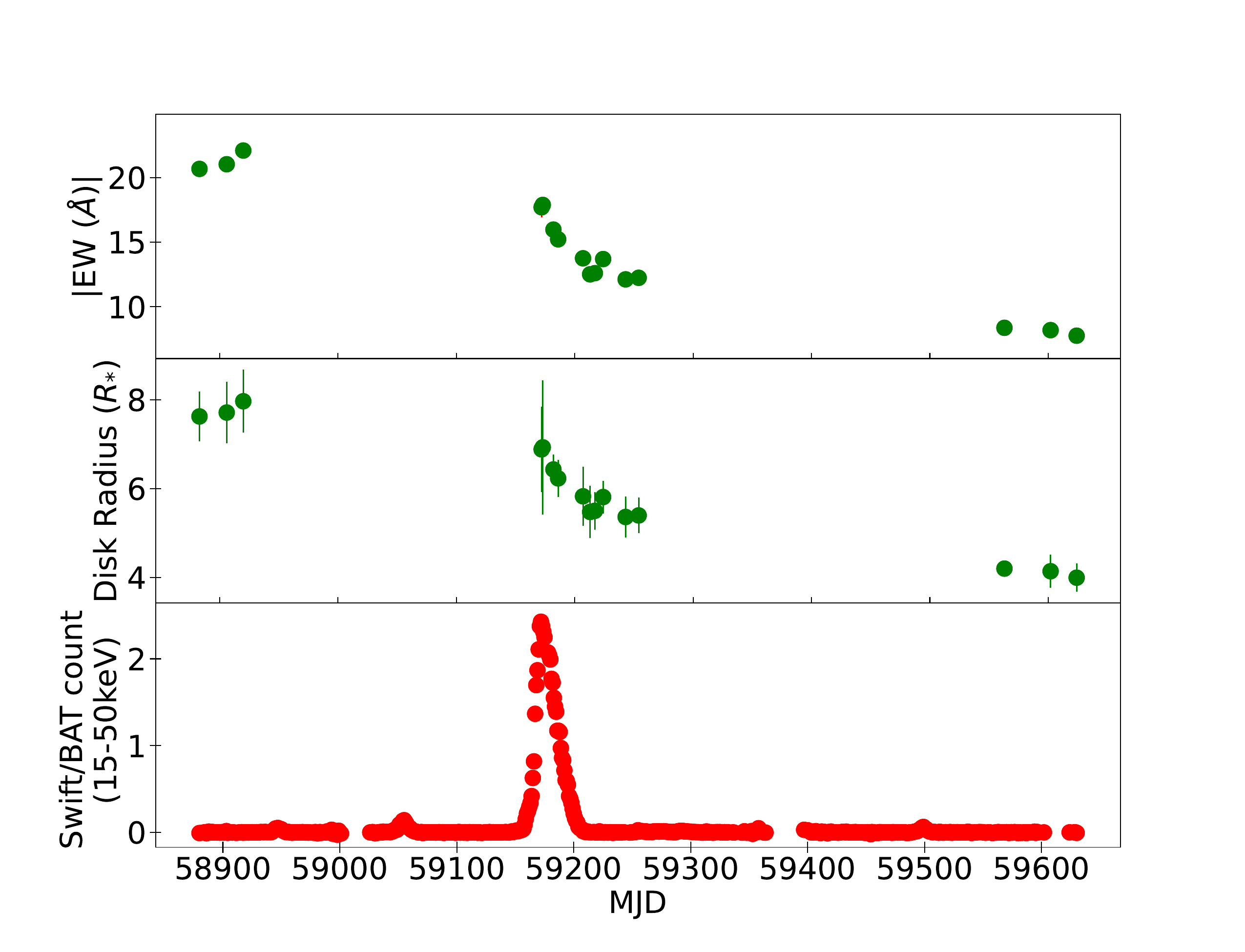}
	\bigskip
	\begin{minipage}{12cm}
	\caption{Variation of the H$\alpha$ line parameters of the Be star and the X-ray count rate from the pulsar in 1A~0535+262/HD~245770 system, during our observations are shown. The changes in line $EW$, circumstellar disk radius, and Swift/BAT X-ray count rate are shown in panels from top to bottom, respectively. The radius ($R_{\ast}$) of the Be star is estimated to be 15 $R_{\odot}$ \citep{okazaki2001natural}.}
	\label{fig:variation_halpha}
	\end{minipage}
\end{figure*}

The equivalent width ($EW$), which signifies the strength of the emission/absorption lines,  is calculated for all observations. We used the area under the curve technique to determine the equivalent width of the lines. The corresponding formula is : {{ $EW$ = $\sum (1- \frac{F(\lambda)}{F_{c}(\lambda)})\delta \lambda$}}, where total flux, F($\lambda$), consists of flux contributions from line and the continuum, while F$_{c}$($\lambda$) represents the underlying continuum flux at wavelength $\lambda$. The errors in $EW$s are calculated using the error propagation method, considering continuum flux and wavelength calibration errors. The chosen waveband for H$\alpha$ lines is 6550-6575 \AA. The continuum flux is determined by taking the median of flux counts in the 6517-6550~\AA~ and 6575-6607~\AA~ wavebands. The obtained $EW$s are plotted in Figure~\ref{fig:variation_halpha}. In absolute terms, the maximum equivalent width of the H$\alpha$ line that was observed during the 2009 and 2011 X-ray outbursts was ($\sim$18 \AA; \citet{article_Moritani}). However, we obtained an |$EW$| of $\sim$22 \AA \hspace{0.01cm} on 2020 March 12.

Most BeXRBs, including 1A~0535+262, exhibit asymmetric H$\alpha$ line profiles. Asymmetries are often described in terms of the V/R ratio, representing the blue-shifted line flux to the red-shifted line flux. Although our observed emission lines are mostly  single-peaked, due to asymmetric nature, we fit all the H$\alpha$ lines with Voigt profiles to determine the contribution of red and blue-shifted flux. Initially, Gaussian functions were attempted for fitting. However, they did not capture the wing regions well. From the fitting, we extracted parameters such as central wavelengths, standard deviations, and intensities of the blue and red components of the H$\alpha$ line. Using this, we can calculate the peak separation in terms of velocity difference $\Delta$$V$, which will help in calculating the disc radius. The H$\alpha$ line parameters can be used to calculate the size of the disc. \citet{1972ApJ...171..549H} proposed that the size of the H$\alpha$ emitting region in the circumstellar disc can be estimated using the peak separation ($\Delta$$V$) of the double-peaked H$\alpha$ emission line, assuming a Keplerian velocity distribution. \citet{1989Ap&SS.161...61H} established a linear relationship between $\Delta$$V$, $V sin i$, and $EW$(in \AA~ unit) for single-peaked profiles. \citet{10.1093/mnras/stw2354} derived the values of the slope ($a$) and intercept ($b$) for this relationship as 0.334 and 0.033, respectively. By applying the equation:
\begin{equation}
	log(\frac{\Delta V}{2V \sin i})= -a \times \log (-EW) + b,
\end{equation}
we calculated the average value of $V~sin i$ to be 236.84$\pm$13.55 km/s, consistent with the reported value of 225$\pm$10 km/s (\citealt{reig2011x} and references therein). The radius of the H$\alpha$ emitting region (R$_{d}$) is calculated using the following equation:
\begin{equation}
	R_{d}= (2V \sin i / \Delta V)^{j} \epsilon R_{\ast} 
\end{equation}
where, j = 2 for Keplerian rotation, $R_{\ast}$ is the radius of the Be companion star (15 $R_{\odot}$ for 1A~0535+262, \citet{okazaki2001natural}), and $\epsilon$ is a dimensionless parameter accounting for various effects that may overestimate the disc radius (typically 0.9$\pm$ 0.1; \citet{10.1093/mnras/stw2354}). The disc radius was estimated during our observations and depicted in the 2nd panel of Figure~\ref{fig:variation_halpha}. Prior to the giant X-ray outburst, the disc size gradually increased until the onset of the outburst, and then steadily decreased thereafter, reaching a value of 4$R_{\ast}$.

\section{Discussion and Conclusion}
We studied the optical and X-ray behavior of 1A~0535+262 during its 2020 giant X-ray outburst using observations from Mnt. Abu Infrared Observatory. We examined the evolution of H$\alpha$ and He~I emission lines in the 1A~0535+262/HD~245770 system before, during, and after the recent giant X-ray outburst in 2020 over a 2-year period to understand the changes in the circumstellar disc. 

Optical spectra of 1A~0535+262/HD~245770 revealed the presence of three emission lines at 6562.8 \AA, 6678 \AA, and 7065 \AA, indicating the presence of a circumstellar disc around the Be star even after the gaint X-ray outburst. These emission lines confirm that the disc did not completely vanish even during and after the giant X-ray outburst \citep{10.1111/j.1365-2966.2006.10127.x}. Previous studies have also observed similar behavior (see, e.g.,   \citealt{10.1046/j.1365-8711.1999.03148.x} for 1A~0535+262 and \citealt{Reig_2014} for IGR~J21343+4738). We examined the evolution of the H$\alpha$ emission line profile before, during, and after the 2020 giant X-ray outburst. The line profiles showed different characteristics in each phase as shown in Figure~\ref{fig:halpha}. Before the outburst, the profile was asymmetric, single-peaked with a broad red wing. During the outburst, it was asymmetric and single-peaked, with the blue-wing broader than the red-wing. After the outburst, the profile remained relatively stable for about a month, with a hump-like structure in the blue wing, and then evolved into a double-peaked shape within a year.

Similar studies were conducted during the 2009 and 2011 giant X-ray outbursts of the 1A~0535+262/HD 245770 system \citep{Camero_Arranz_2012}, where the H$\alpha$  line profiles were found to be marginally asymmetric and single-peaked. The profile during the 2009 outburst had a broader blue wing, while the 2011 outburst showed a broader red wing. Comparing the H$\alpha$  line profile before the 2020 outburst to the 2009 outburst, we observed similarities, with the post-outburst profiles being red-shifted with a broader blue wing, while the pre-outburst profiles were blue-shifted with a broader red wing. Higher spectral resolution (R$\sim$ 30000-60000) by \citet{article_Moritani} around the 2009 outburst indicated that the basic structure of the emission line is consistent with our findings and those from previous studies \citep{Camero_Arranz_2012}. \citet{article_Moritani} also reported several absorption features in the H$\alpha$ line profile suggesting a complex structure of the Be circumstellar disc. From a long term observation of 4U~0115+634/V635~Cas, \citet{2001A&A...369..108N} and \citet{2007A&A...462.1081R} suggested that a precessing warped Be circumstellar disc caused the change in the profile from a normal double-peaked to a single-peaked profile during the giant X-ray outburst. Similar behaviour was also observed by \citet{10.1093/pasj/63.4.L25} during the 2009 giant outburst in 1A~0535+262. 

In our study, we observed a maximum |$EW$| of H$\alpha$ emission line ($\sim$22 \AA) before the 2020 giant X-ray outburst, in comparison to that observed prior to the 2009 X-ray outburst ($\sim$18 \AA; \citet{article_Moritani}). Hence, it is evident that a larger Be circumstellar disc was formed prior to the 2020 outburst, before being consumed by the neutron star. One plausible explanation for this observation could be the longer timescale required for the tidally warped region to descend towards the path of the neutron star in the binary plane. This prolonged descent may have allowed for the development of an extremely large disc. The measured Be disc size was largest before the outburst, but decreased afterwards. The disc radius before the outburst was $\sim 8.27 \times 10^{10}$ m, equivalent to 7.62 times the star radius. The calculated 4:1 truncation radius was  $7.27\times 10^{10}$ m, suggesting the disc size is limited by viscous and tidal torque balance \citep{2001A&A...369..108N}. According to  \citet{okazaki2007interaction}, for highly eccentric and misaligned (spin axis of the Be star and the normal to the binary plane are not parallel) viscous decretion disc, the truncation is less efficient when the inclination ($i$) is $\geq$ 60$^{\circ}$. This is indicating a highly misaligned disc. The highly misaligned disc with respect to the binary plane could be warped \citep{10.1111/j.1365-2966.2011.19231.x,10.1093/pasj/65.2.41}, allowing the neutron star to capture more matter during the outburst. Similar warping was observed in 2009 \citep{10.1093/pasj/65.2.41}. 

The torque from the Be star is crucial for disc warping. When the Be star expels matter, it aligns the disc with its equatorial plane. However, in the absence of matter supply, the disc can easily warp towards the orbital plane due to tidal interaction of neutron star. This lack of matter reduces the V-band flux from the inner region of the disc. Previous  studies by \citet{Camero_Arranz_2012} have shown giant X-ray outbursts during the optical fading phase in 1A~0535+262. \citet{Yan_2011} also found an anti-correlation between the $V$-band magnitude and the $EW$ of the H$\alpha$ line prior to the 2009 outburst which is an indication of mass loss from the inner part of the Be disc. An increase in $V$-magnitude before the 2020 outburst (see Figure~9 of \citet{2023MNRAS.518.5089C}) suggests mass ejection from the Be star, resulting in dimming the continuum emission. This material likely contributes to the growing circumstellar disc. As a result, we expect a strong H$\alpha$ line before the outburst as shown in Figure~\ref{fig:variation_halpha}. The large disc size leads to a strong X-ray outburst. After the outburst, the neutron star truncates the circumstellar disc, causing a rapid decrease in H$\alpha$ emission.

In addition to that, to understand evolution of the inner disc due to outburst, we analyzed the H$\alpha$ line's Full Width at Zero Intensity (FWZI). The FWZI indicates the velocity components of the inner disc region. In a scenario without matter supply from the Be star, the density in the inner disc decreases, resulting in a reduction of the high-velocity component of the H$\alpha$ line (FWZI). Comparing the pre-outburst and peak outburst observations, we found the average FWZI of the H$\alpha$ line to be approximately 1017 km/s before the outburst, while during the peak, it was estimated to be around 771 km/s. This suggests that the inner part of the disc was absent before the giant outburst, which affected the warping of the outer part of the disc and potentially triggered the neutron star's giant X-ray outburst.

In conclusion, our study of the optical properties of BeXRB 1A~0535+262/HD~245770 during the 2020 giant X-ray outburst revealed interesting findings. Prior to the outburst, the H$\alpha$ emission lines displayed asymmetric and a single-peaked profile with a broad red wing. However, during the outburst, the line profiles exhibited the opposite pattern with a broad blue wing. After approximately 400 days, the profiles transitioned to a double peak. The increasing $EW$ of the H$\alpha$ line before the outburst suggested a larger disc size, exceeding the 4:1 resonance radius of the decretion disc, indicating a highly misaligned disc. The transformation from a single to a double peak profile and the presence of a misaligned disc suggest that mass accretion from a warped disc onto the neutron star triggered the giant outburst.

\begin{acknowledgments}
We thank the anonymous reviewer for his/her comments on the paper. The authors thank Dr. Mudit K. Srivastava and all other MFOSC-P instrument team members of Physical Research Laboratory, India, for their constant support during the optical observations at various epochs. The research work at Physical Research Laboratory (PRL) is funded by the Department of Space, Government of India.
\end{acknowledgments}

\begin{furtherinformation}

\begin{orcids}
\orcid{0000-0003-2865-4666} {Sachindra} {Naik}
\orcid{0000-0002-9680-7233} {Birendra} {Chhotaray}
\orcid{0000-0003-0071-8947} {Neeraj} {Kumari}
\end{orcids}

\section*{Author contribution}
Conceptualization, S.N. and B.C.; Formal Analysis, B.C.; Methodology, S.N.; Investigation, B.C.; Resources, N.K., S.N., and B.C.; Writing- Original Draft, S.N. and B.C.; Writing-review and editing, S.N., and N.K.; Supervision, S.N.

\section*{Conflicts of interest}
The authors declare no conflict of interest.
\end{furtherinformation}
\bibliographystyle{bullsrsl-en}

\bibliography{extra}

\end{document}